\documentclass[a4paper,fleqn,usenatbib]{mnras}
\usepackage{mathptmx}
\usepackage[T1]{fontenc}
\usepackage{ae,aecompl}
\usepackage{graphicx}
\usepackage{amsmath}
\usepackage{amssymb}

\title[Search for exoplanets and brown dwarfs]{Search for exoplanets and brown dwarfs with VLBI.}
\author[K. Katrzy\'nski et al.]{
K. Katarzy\'nski$^{1}$\thanks{E-mail:kat@astro.uni.torun.pl}
M. Gawro\'nski$^{1}$
K. Go\'zdziewski$^{1}$\\
$^{1}$Centre for Astronomy, Faculty of Physics, Astronomy and Informatics,\\
Nicolaus Copernicus University, Grudziadzka 5, PL-87-100 Torun, Poland
}
\begin{document}

\date{Accepted ... . Received 2015 February 20; in original form 2015 January 2}

\pagerange{\pageref{firstpage}--\pageref{lastpage}} \pubyear{2016}

\maketitle

\label{firstpage}

\begin{abstract}
   The main aim of this work is to estimate possible radio GHz emission of extrasolar planets and brown 
   dwarfs and to check if such radiation can be detected by Very Large Baseline Interferometers (VLBI).
   In the estimation we assume that the emission may originate in processes similar to those observed 
   in the Jupiter system. The frequency of the radio emission that is produced in this system depends 
   mostly on the magnetic field strength. Jupiter's magnetic field ($\sim 9$ G on average) allows for 
   radiation from kHz frequencies up to 40 MHz. This is well below the frequency range of VLBI. However, 
   it was demonstrated that the magnetic field strength in massive and young object may be up to two orders 
   of magnitude higher than for Jupiter, which is especially relevant for planets around short-lived A type stars.
   This should extend the range of the emission up to GHz frequencies. We calculated expected flux densities of 
   radio emission for a variety of hypothetical young planetary systems. We analysed two different emission scenarios, and found that the radiation induced 
   by moons (process similar to Jupiter-Io interactions) appears to be less efficient than the emission 
   generated by a stellar wind on a planetary magnetosphere. We also estimated hypothetical emission of 
   planets and brown dwarfs located around relatively young and massive main sequence A-type stars. Our 
   results show that the emission produced by stellar winds could be detected by currently operating VLBI networks.
\end{abstract}


\section{Introduction}

\begin{figure}
\centering
\includegraphics[width=\columnwidth]{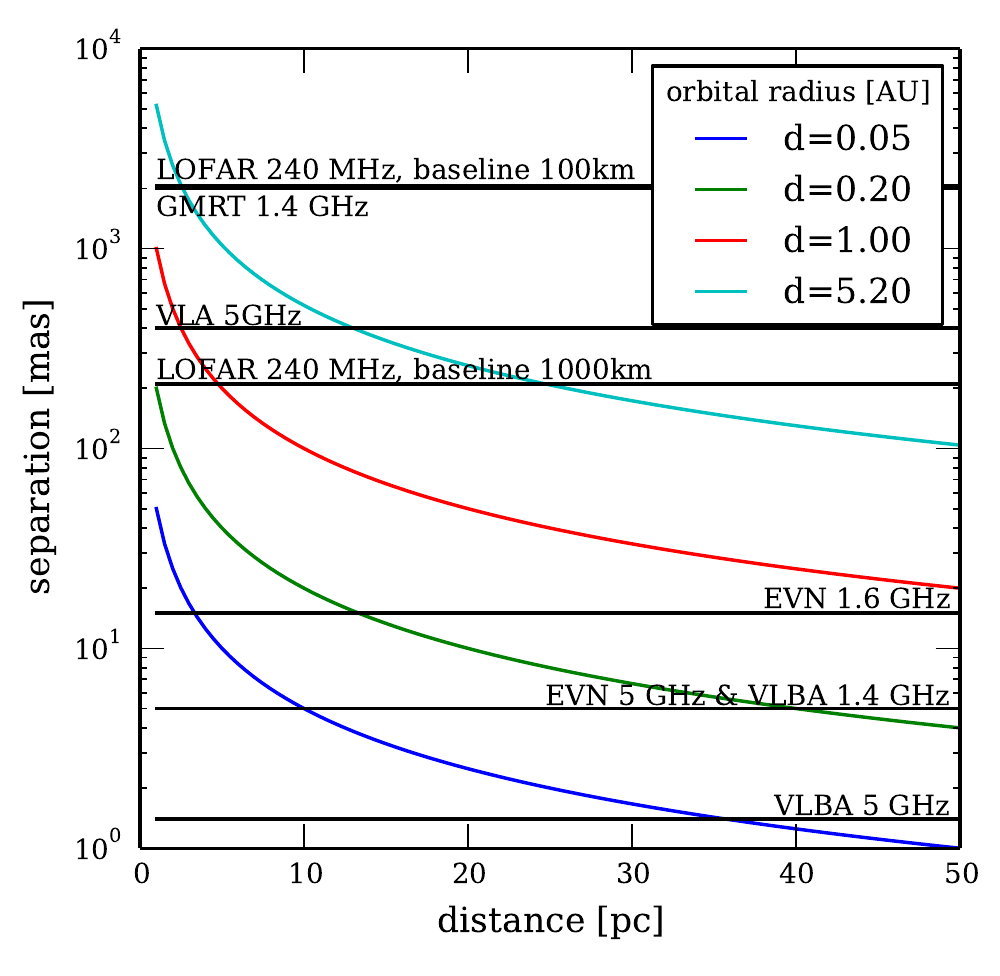}
\caption{
The expected maximal separation between stars and planets estimated for
different orbits and distances (curved lines). Horizontal lines with labels 
show the angular resolution of the main interferometric networks in the 
frequency range from 0.24 to 5 GHz. The angular resolutions are gathered 
from the specifications provided at the web pages of the projects.
}
\label{fig_0}
\end{figure}

The angular resolution of very large baseline radio interferometers like the Very Long 
Baseline Array (VLBA\footnote{\url{www.vlba.nrao.edu}}) or the European VLBI Network 
(EVN\footnote{\url{www.evlbi.org}}) can reach a few milliarcseconds
(note that the resolution depends on the frequency of observations, see Fig. \ref{fig_0}). 
This may give an opportunity for direct observations of extrasolar planets in nearby stellar 
systems. However, the main problem  of detecting low mass companions in these systems is the 
frequency of the observations. All radio telescopes in such interferometric networks 
are parabolic antennas, designed to work efficiently at GHz frequencies ($\nu \ge 1.2 
\; \rm GHz$ for the VLBA and $\nu \ge 1.4 \rm \; GHz$ for the EVN). On the other hand, 
observations of planets in the Solar system and especially observations of 
the Jupiter system show that the dominant 
radio emission extends there from kHz frequencies up to 40 MHz. If this 
emission range is typical for all massive extrasolar planets then their radiation 
could be likely detected only by a low frequency interferometers like the Low Frequency 
Array for Radio Astronomy (LOFAR\footnote{\url{www.lofar.org}}). If the emission
extends up to a few hundred MHz. Then it should be possibly observed by the Giant 
Metrewave Radio Telescope (GMRT\footnote{\url{gmrt.ncra.tifr.res.in}}). However, 
such interferometers have limited angular resolution due to significantly smaller 
baselines and relatively low frequency of observations, when compared to VLBI 
networks.

The theoretical models of the MHz radio emission from massive exoplanets 
were already investigated by several authors \citep[e.g.][]{Griessmeier05, 
Griessmeier07, Reiners10, Nichols11, Noyola14}, for a review see \citet{Griessmeier11}. 
It was demonstrated that in principle, 
it should be possible to detect such emission from at least a few objects (e.g., 
$\tau$ Boo b, Gl 86 b, GJ 3021 b, eps Eridani b, Gliese 876 b). However, 
despite many observational trials  \citep[e.g.,][]{Winglee86, Bastian00, George07, 
Lazio07, Lazio10, Lecavelier13} there is no confirmed detection that could be addressed 
as a radio emission of an exoplanet. Note, that recently \citet{Sirothia14} found 
sources of radio emission towards four planetary systems. However, these tentative
detections require further investigation because the emission may come from coincidental 
background sources.

In this work we explore higher frequency range to check if the
radio emission from massive planets can be observed at GHz frequencies with 
the VLBA or the EVN interferometers. This frequency range was not well
explored from the theoretical and the observational point of view 
\citep[e.g.,][]{Griessmeier2006, Shiratori2006}. On the other hand the radio
interferometry in the GHz range provide very good angular resolution, and
potentially can be very useful for the search of exoplanets.
In Fig. \ref{fig_0} we compare the angular resolution of the main
interferometric networks with the angular separation for different 
planetary systems, observed from different distances. The comparison shows 
that the angular resolution of the interferometers that work at GHz frequencies
can be one or even two orders of magnitude better than the resolution of 
the instruments that operate in the MHz range. Note that according to the 
estimations presented in this paper we should not expect any emission around
5 GHz. Thus, the angular resolutions at 5 GHz in Fig.~\ref{fig_0} are
plotted just for comparison. If we focus at 1.4 GHz, possible planets 
could be resolved from the distance smaller than 10~pc for the orbital 
radius of about 0.05 au. However, there are only four interesting stars within the 
radius of 10 pc. Therefore, the reasonable distance between the star and 
the possible planet, that we consider here, is about 1 au.

Interferometers like EVN or VLBA provide also an excellent sensitivity. 
For example, after 2h 
of integration with the recording rate 1 Gb/s, the EVN is able  to reach $3\sigma$ 
detection level for sources with fluxes $\sim 35 \; \mu \rm Jy$ at 1.4 GHz.
In  Table \ref{tab_1} we specify sensitivities for main VLBI networks at different
frequencies and recording rates. Note that an increase of the recording rate extends
the bandwidth (for example 1 $\to$ 2 Gb gives 128 $\to$ 256 MHz). The sensitivities
presented in Table \ref{tab_1} are given per beam. However, for point like
sources, which is the case of the radio emission expected from planets, these
sensitivities are equivalent to values given in Jy.

The sensitivity of EVN at 1.4 GHz is three orders of magnitude 
better than the sensitivity of GMRT at 150~MHz for comparable integration time 
\citep[e.g.,][]{Sirothia14}. The comparison of the sensitivities at different
frequencies can be misleading. Especially if the observed spectra are steep 
($F \propto \nu^{-\alpha}$, $\alpha \gg 1 $) or if such spectra contain an abrupt 
cut-off above some maximum MHz frequency. In general, the spectrum detectable at GHz 
frequencies should also be observed at MHz frequencies. However, if there is no 
significant increase of the MHz emission, this region of the spectrum can be 
problematic for observations, because of relatively low sensitivity of LOFAR or 
GMRT. This illustrates the potential of radio interferometers that contain big antennas 
and shows that the VLBI could potentially provide very important observations 
of exoplanets. However, the question is if we can expect any radio emission from 
exoplanets at the GHz frequencies?

\begin{table}
\begin{center}
\begin{tabular}{lcrrr}
$\frac{\rm Network}{\rm name}$  & $\frac{\rm Recording}{\rm rate}$(Gb\:s$^{-1}$) & $\frac{\rm 1.4 \; GHz}{\rm \mu Jy/beam}$ & $\frac{\rm 1.6 \; GHz}{\rm \mu Jy/beam}$ & $\frac{\rm 5 \; GHz}{\rm \mu Jy/beam}$  \\ 
\hline
VLBA  & 2 & 22.7 & 24.7  & 16.5 \\
e-EVN & 2 & 10.3 &  9.5  &  9.6 \\
e-EVN & 1 & 14.6 & 13.4  & 13.6 \\
EVN   & 1 & 11.9 & 10.4  & 10.7 \\
EVN   & 2 &  7.9 &  7.3  &  7.6 \\
\hline
\end{tabular}
\end{center}
\label{tab_1}
\caption{Levels of thermal noise ($1\sigma$) expected on interferometric images for the main VLBI 
networks. The values were obtained using the EVN Calculator (\url{www.evlbi.org/cgi-bin/EVNcalc}),
where we assumed 2 h of the integration time.
}
\end{table}

The main motivation for the calculations presented in this work is
the estimation of the magnetic field strength made by \citet{Reiners10}.
They demonstrated that in young (age $< 10^8$ yr) and massive
(from a few to several Jupiter masses) objects the magnetic
fields strength may even excess 1~kG. This is two orders of 
magnitude larger than the value of polar dipole magnetic field 
strength of Jupiter ($\sim11\rm \; G$ south pole , $\sim14\rm \; G$  north pole). 
The maximum frequency of the emitted radiation is simply the cyclotron 
frequency, that is directly proportional to the magnetic field strength. 
Therefore, the emission that is observed in the Jupiter system
below 40 MHz, in younger systems could be 
generated at GHz frequencies. It should be also mentioned that
the radio emission at GHz frequencies has been discovered
already in ultra cool dwarfs of L, M spectral types \citep[e.g.,][]{Berger01, 
Hallinan07, Hallinan08, McLean12} and also in a T6.5 spectral type 
brown dwarf J1047+21 \citep{Route12, Route13}. The mechanism of such emission is probably 
similar to the radiation observed in the Jupiter system. This is a coherent 
emission powered by the electron cyclotron maser instability 
\citep[e.g.,][]{Treumann06}. The emission observed from J1047+21 
is sporadic and has a form of short burst (duration from tens to 
hundreds of seconds). Short spikes observed during
the burst allows to constrain the size of the emitting region to
$\leq 0.4 R_{\rm J}$. The independent estimation derived from the 
frequency drift during the burst, give the size in the range $0.3-1 R_{\rm J}$. 
Nevertheless, the fact that the emission of J1047+21 was observed around 
4.75 GHz indicates that the magnetic field strength in this particular 
object is $B \simeq 1.7 \; \rm kG$. This value is similar to the strengths 
obtained for L, M type dwarfs. The magnetic field strength estimated
in a case of ultracool dwarf TVLM 513-46546 (spectral type M9) was 
around 7 kG \citep{Yu2011}. The observational findings described above 
give additional motivation to check what exactly we can expect observing 
massive planets using the VLBI at GHz frequencies.

Among a few possible processes that can generate radio emission 
in the planetary systems (see for a review \citeauthor{Griessmeier11}
\citeyear{Griessmeier11}), we focus on two possible scenarios, that are 
well known from the Solar system. The first mechanism assumes an interaction 
between the stellar wind and the planetary magnetosphere. The power of the radio 
emission in such process is proportional to the kinetic energy flux of 
particles that are impacting on the planet magnetopause \citep[e.g.,][]{Zarka97,
Farrell99, Stevens05}. Such process is responsible for the most of 
the Jupiter's hectomeric (HOM) emission. The second scenario assumes the 
existence of a moon around a planet. This is mechanism similar to the process
that generates the Io decamertic (DAM) emission in the Jupiter system. The 
volcanic activity of Io fills the magnetosphere with matter 
(sulphur, oxygen) that is ionized and accelerated by electric currents
inducted by Io. The currents are inducted because of the difference
between the Jupiter's rotation velocity and the Io's orbital velocity
\citep[e.g.,][]{Nichols11, Nichols12, Noyola14}. 

It must be mentioned that there is also a third important emission scenario. 
According to this scenario the radiation is produced by the interaction between 
the magnetic energy flux of the interplanetary magnetic field with the planetary
magnetosphere \citep{Zarka01, Farrell04, Griessmeier07}. In the Solar system
it is impossible to distinguish which emission process dominates, this one or the 
dissipation of the wind kinetic energy. However, in this work we are going to focus 
on main sequence A-type stars, were the wind velocity, the key parameter
for the first mentioned emission scenario is a few times higher than in the
Solar system. On the other hand the third process depends strongly on the interplanetary magnetic 
field that is difficult to estimate or measure in the case of main sequence 
A-type stars. Some observational evidences \citep[e.g.][]{Lignieres2009, Blazere2016}
suggest that magnetic field of such stars is similar or less ($B \lesssim 1$ G) 
to the value observed for the quiet Sun. This may favour our first emission
scenario over the third process discussed here. However, this requires further
detailed investigation and is out of the scope of this paper.  

The most important parameter for all emission scenarios discussed above
is the magnetic field strength. There are several different
models that describe how the magnetic field of planets and brown
dwarfs can be generated \citep[for a review see][]{Christensen10}. 
In the next section we describe the model selected for our calculations 
and discuss this particular choice. The description  of first two mentioned 
above emission scenarios is given in Sections 3 and 4. In the Section 
5 we compare fluxes expected from these scenarios, calculated for a wide range of ages and 
masses of hypothetical planetary systems. Finally, we focus on selected A-type 
main sequence stars, where planetary systems could possibly be observed due to 
their young age and strong magnetic fields. Note that recently \citet{Nielsen2013}
reported results of the Gemini NICI Planet-Finding Campaign. They conducted direct
imaging of 70 young B and A--type stars, searching for planets and brown
dwarfs. As the result they identified two new low-mass companions to HD 1160 and 
HIP 79797 stars. They also investigated previously discovered planet $\beta$
Pic b \citep{Lagrange09}, estimating their orbital parameters. They found for
example the semi-major axis of this object to be in the range 8.2-48~au with
95\% confidence \citep{Nielsen2014}. On the other hand the radio interferometry
may potentially provide significantly better angular resolution. According to 
the parameters plotted in our Fig. \ref{fig_0}, at the distance of about 20~pc 
(distance to $\beta$ Pic) radio interferometers should resolve planets with 
the orbital radius less than 1 au. Thus, the main motivation of this work is to
estimate, if the radio interferometry at GHz frequencies can be useful for the search
of exoplanets and brown dwarfs.

\section{Magnetic field estimation}

\begin{figure}
\centering
\includegraphics[width=\columnwidth]{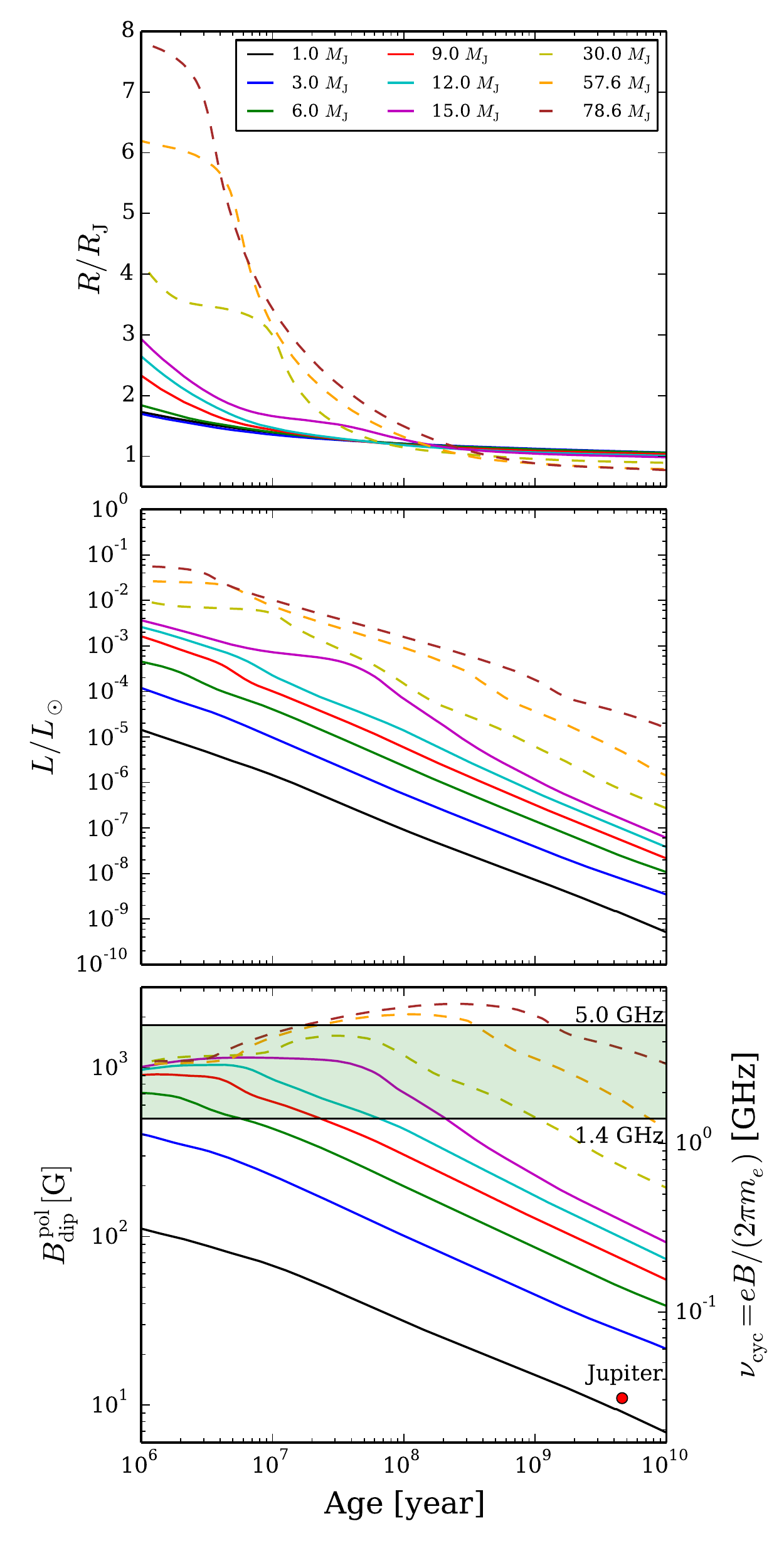}
\caption{The evolution of radius and luminosity calculated for 
different masses of giant planets ($\la 15 M_{\rm J}$, solid lines) 
and brown dwarfs ($\la 80 M_{\rm J}$, dashed lines), according to the model
by \citet{Burrows93, Burrows97}. The lower panel shows estimated  
value of polar dipole magnetic field strength and the corresponding
cyclotron frequency. Jupiter's magnetic field strength, that
is about 2 G higher than the prediction of the model, is indicated by
a dot. A typical range of frequencies used by the global
interferometers is indicated by the shaded area.
}
\label{fig_1}
\end{figure}

The magnetic field observed on the surface of planets in the Solar system 
originates from the dynamo mechanism. The field is inducted by circulating 
electrical currents created inside a fluid interior of a planet. The
circulation comes from the Coriolis force and the convecting flows
supported by the internal heating. Therefore, the strength of the
field may depends on many parameters. The most important of them are 
the density, the conductivity and the size of electrically conducting 
fluid core of the planet. Important are also the convected energy
flux in the core and the rotation rate. Several simple but completely 
different relations were proposed to connect these mostly unknown 
parameters with observed values of the magnetic fields \citep[e.g.,][]{Christensen10}. 
Such relations are called dynamo scaling laws. The diversity of the
proposed solutions is confusing, especially because all of them 
provide values of the magnetic field strength that are in a good agreement
with the observations made in the Solar system. Therefore, it is
difficult do decide which solution should be used, and especially 
which approach should be adopted for extrasolar planets.

In this work we decided to adopt the scaling law given by \citet{Reiners09}
and the approach proposed by \citet{Reiners10}, that was based on the work of \citet{Christensen09}. 
According to this scaling law the dynamo magnetic field 
strength at the surface of a planet is given by
\begin{equation}
B_{\rm dyn} = 4.8 \times 10^3 \left(\frac{M L^2}{R^7} \right)^{1/6} (\rm Gauss),
\label{equ_1}
\end{equation}
where mass ($M$), luminosity ($L$) and radius ($R$) of the planet are given in 
the solar values. We may derive the parameters required by the above formula 
from observations or from theoretical models. This illustrates advantage of this
relationship in comparison to the other scaling laws that usually require the density,
the conductivity and the size of the fluid core, the parameters that are difficult to estimate.
Moreover, this formula (in opposition to the other scaling laws) does 
not depend on the rotation rate, which also is difficult to estimate.
It is enough if the rotation velocity is higher 
than the critical velocity, what should be true for most of massive 
planets and brown dwarfs \citep{Reiners08, Christensen09}. Finally,
the dynamo scaling law described by equation (\ref{equ_1}) predict that the 
magnetic field strength in young massive planets and brown dwarfs may 
excess 1 kG. This is of crucial importance for our calculations. It is necessary to
reach this level of the magnetic strength  in order to expect any 
emission at GHz frequencies. Therefore our results strongly depend 
on the selected scaling law. If this law is not accurate or cannot 
be used to some of object, then our predictions should be revised.

The dynamo magnetic field is related to the polar dipole magnetic 
field strength by a simple formula
\begin{equation}
B_{\rm dip}^{\rm pol} = \frac{B_{\rm dyn}}{\sqrt{2}} \left(1 - \frac{0.17}{M/M_{\rm J}}\right)^3,
\end{equation}
where $M_{\rm J}$ is the Jupiter mass. It is assumed here that the
polar magnetic field strength is two times larger than the equatorial
field strength. The parameters required in the first equation were
obtained from the evolutionary models proposed by \citep{Burrows93, 
Burrows97}. Those relatively old models confronted with the recent
observations \citep{Burrows11} and calculations \citep{Marleau14} 
appears to be precise enough for our estimations. 

In Fig. \ref{fig_1} we show the evolution of the main physical parameters 
($R, \; L$) and estimated magnetic field strength for objects
with different masses. The upper panel in this figure demonstrates 
that above the age of about $10^8$ yr, objects with different masses
have similar radii ($\sim 1 R_{\rm J}$). Therefore, this parameter
is of less importance in the estimations, starting from this age. The middle panel
demonstrates the evolution of the luminosity in comparison to the solar luminosity. 
This parameter depends on the mass but decreases in time, in different object, in a similar way.
Note the discrepancy in luminosity of about four orders of magnitude for objects
with different masses. In the lower panel we show estimated $B_{\rm dip}^{\rm pol}$ 
and the corresponding cyclotron frequency. The magnetic field strength 
from a few hundreds Gausses up to values above 1 kG should be expected 
only in relatively young giant planets (age $\la 4 \times 10^7$ yr).  
Whereas in brown  dwarfs we can expect a strong magnetic field also in old 
objects (age $\ga 10^9$ yr). This means that the emission
at GHz frequencies should be detectable in many brown dwarfs.
This is already confirmed by the detection that we have quoted 
above.

\section{Interaction with the wind}
\label{sec_wind}

The main aim of this work is to provide simple estimations of the
expected radio emission from different planetary system. Since most
of physical parameters in such systems are usually unknown, we use
as simple as possible description of the emission processes, where
most of the parameters can be derived or extrapolated from the 
Solar system.

First, we analyse the interaction of a stellar wind with the planet 
magnetosphere. This process is similar in origin to the Jupiter's 
HOM radiation. To calculate the total radiated power we adapted a simple 
formula derived by \citet{Griessmeier05}
\begin{equation}
P_1 = \left(\frac{\mathcal{M}}{\mathcal{M}_{\rm J}}\right)^{2/3}
    \left(\frac{n}{n_{\rm 1au}}\right)^{2/3}
    \left(\frac{v}{v_{\rm 1au}}\right)^{7/3}
    \left(\frac{d}{d_{\rm J}}\right)^{-4/3}
    P_{\rm J},
\label{equ_P1}    
\end{equation}
where the magnetic moment is given by
\begin{equation}
\mathcal{M} = 4 \pi \frac{B_{\rm dip}^{\rm pol} R^3}{2 \mu_0},
\end{equation}
where $n$ is the wind particle number density, $v$ is the wind 
velocity and $d$ is the planet to star distance. All these parameters are
normalized by the Solar system values ($\mathcal{M_{\rm J}} = 1.5 \times 10^{27} \; \rm A m^2$,
$n_{\rm 1au} = 6.59 \times 10^6 \; \rm m^{-3}$, $v_{\rm 1au} = 425 \; \rm km/s$,
$d_{\rm J} = 5.2 \; \rm au$). This formula depends on the 
assumed power of the emission in the Jupiter's system $P_{\rm J}$.
The average value of $P_{\rm J}$ is of about $3.1 \times 10^{10} \; 
\rm W$. However, the average power during high activity periods is
$P_{\rm J} = 2.1 \times 10^{11} \; \rm W$ and the peak power may
reach even $P_{\rm J} = 1.1 \times 10^{12} \; \rm W$ \citep{Zarka04}. 
This gives the discrepancy reaching two orders of magnitude. In our estimations we
use the average power during high activity periods. This means
that we are expecting observations in a preferable conditions.
Therefore, the estimated flux should be treated as the maximum possible 
value.

The crucial factors in equation (\ref{equ_P1}) are the wind parameters.
In principle the velocity and the density of wind we can derive from
an appropriate wind model. In the case of main sequence G, K, M type
stars winds are mostly driven by gas pressure gradients in the corona
\citep{Parker58} and various additional acceleration processes 
\citep[see for a review][]{Echim11}. Parker's wind model was 
used for example by \citet{Griessmeier07} to predict low--frequency 
radio fluxes of known extrasolar planets. 

Here, we are going to focus on
main sequence A-type stars. Such object are much more luminous than G, K, M 
stars (by 1-2 orders of magnitude). Therefore, we assume that
the winds of such stars are driven by the radiation pressure.
 The radiative force that accelerates the wind comes 
from the scattering on free electrons and interception of
photons by ions of the atmospheric plasma. Ions in turn produce
the observed spectral lines. Thus such winds are frequently 
called line--driven winds. First models of such winds
were proposed by \citet{Lucy70} and \citet{Castor75}, to explain 
mass loss rates in O-type stars. In last few decades the models
were successively improved and today we may speak about a family 
of CAK models (after Castor, Abbott and Klein 1975). In should
be mentioned that  the CAK theory was successfully applied to O, B 
type stars \citep[e.g.,][]{Vink00, Kudritzki02, Puls08} and A, F, G 
supergiants \citep[e.g.,][]{Achmad97, Cure11}. There are only
a few works that try to apply the CAK theory to main sequence
A-type stars \citep[e.g.,][]{Babel95, Bertin95, Vick10}. This
may be related to the fact that there are no direct observational
measurements for mass loss rates in main sequence A-type stars.
The upper limits derived from the observations \citep{Brown90, Lanz92}
are 2-3 orders of magnitude above very few estimations
\citep{Bruhweiler91, Bertin95} we have for such stars.

According to the standard CAK theory the mass loss rate of a star 
can be approximated by
\begin{equation}
\dot{M}_{\rm CAK} \approx \frac{L_*}{c^2} \frac{\alpha}{1-\alpha} \left(\frac{Q \Gamma}{1-\Gamma} \right)^{\frac{1-\alpha}{\alpha}},
\label{equ_mass_loss}
\end{equation}
\citep[e.g.,][]{Owocki04} where $L_*$ is the star bolometric luminosity, 
$\alpha$ is one from three so-called line force multiplier parameters 
(assumed here to be 0.5) and
\begin{equation}
\Gamma = \frac{a_r}{a_g} = \frac{\sigma_e L_*}{4 \pi c G M_*} 
\end{equation}
is the Eddington factor that relates the radiative acceleration by
the scattering on free electrons ($a_r$) with the gravitational 
acceleration ($a_g$). Other parameters in the above formula
are mass of a star $M_*$, the electron scattering opacity
$\sigma_e = 0.325$ cm$^2$g$^{-1}$, and the speed of light $c$. 
Note that the mass loss rate given
by equation (\ref{equ_mass_loss}) was derived under assumption that
the star was a point source at the origin. Improved implementations 
of the CAK theory \citep[e.g.,][]{Friend86, Pauldrach86} takes into 
account the finite size of the star and the centrifugal force 
due to the star's rotation. This reduces the mass loss rate by
a factor
\begin{equation}
\dot{M} \approx \frac{\dot{M}_{\rm CAK}}{(1+\alpha)^{1/\alpha}}.
\end {equation}
Equation (\ref{equ_mass_loss}) contains also the dimensionless line 
strength parameter $Q$ that replaced two other force multiplier 
parameters, referred in the CAK terminology as $k$ and $\delta$
\citep[e.g.,][]{Abbott82, Owocki04}. The $Q$ parameter depends
on the star metallicity. In the case of O, B stars $ Q \simeq 10^5 Z$,
with the assumption that $Z_\odot = 0.019$ \citep{Gayley95}. However,
this parametrization of Q used for A-type stars gives mass loss 
rates order of magnitude higher than expected values. We
verified our calculations with the mass loss rates estimated for
the Sirius A star. An early findings by \citet{Bertin95} based 
on the observations of Mg II lines give for this object 
$2 \times 10^{-13} < \dot{M} <  1.5 \times 10^{-12} \; \rm M_\odot/yr$.
However, more recent investigations suggest the mass loss rate in
Sirius A should be of the order of $10^{-13} \; \rm M_\odot/yr$ or 
less. A higher mass loss rate ($\sim 10^{-12} \; \rm M_\odot/yr$) would 
not allow to reproduce observed surface abundance patters for
this star \citep{Vick10}. Assuming $ Q = 1.5 \times 10^4 Z$, we
obtained ($2 \times  10^{-13} \; \rm M_\odot/yr$) for Sirius A. This 
assumption is also in a good agreement with the mass loss rate
estimated for $\beta \; \rm Pic$ $\dot{M} \simeq 1.1 \times 10^{-14} 
\; \rm M_\odot/yr$ \citep{Bruhweiler91}, where our calculation
gives $\dot{M} = 10^{-14} \; \rm M_\odot/yr$.

The CAK wind velocity law is given by
\begin{equation}
v_w(d) = v_\infty \left(1 - \frac{R_*}{d}\right)^\beta,
\label{equ_wind_vel}
\end{equation}
where $R_*$ is the star radius, $\beta$ describes the velocity 
profile and lies in the range $0.5 < \beta < 1$. In the standard
CAK theory (point--like star) $\beta = 0.5$. Improved CAK models
suggest $\beta \approx 0.8$, the value that we used in this 
work. However, detailed value of $\beta$ is not important for
$d \gg R_*$, when the wind velocity becomes equivalent to
the terminal velocity, given by
\begin{equation}
v_\infty \approx 2.25 \frac{\alpha}{1-\alpha} v_{\rm esc},
\end{equation}
that is of the order of the effective escape velocity
\begin{equation}
v_{\rm esc} = \sqrt{\frac{2 G M_{\rm eff}}{R_*}},
\end{equation}
where the effective mass $M_{\rm eff} = M_*(1-\Gamma)$ combines the radiative 
acceleration on electrons and the gravity. The wind velocity obtained from equation (\ref{equ_wind_vel} )
must be transformed to the reference frame of the planet
$v = \sqrt{v^2_w + v^2_{\rm K}}$, where $v_{\rm K}$ is the Keplerian velocity
of the planet. However, this effect is important only for planets at relatively
tight orbits, where the Keplerian velocity is comparable to the wind 
velocity.

Having the mass--loss ratio and  the wind velocity, we may easily obtain the
particle number density using the mass continuity equation
\begin{equation}
\dot{M} = 4 \pi d^2 n(d) v(d) m_{\rm p},
\end{equation}
where $m_{\rm p}$ is the proton mass.

\section{Interaction with a moon}
\label{sec_moon}

The second emission scenario assumes the existence
of a moon around the planet. This is a mechanism similar to the Io-Jupiter
interactions that produce the so-called Io-DAM emission. We follow here the
approach proposed by \citet{Noyola14}, where the formula for the
maximum Joule dissipation in the Jupiter system \citep{Neubauer80}
was used. According to this formula, the power of radio emission
related to the planet--moon interaction is given by
\begin{equation}
P_2 = \beta \frac{\pi R^2_{\rm m} V^2 B^2_{\rm m}}{\mu_0\sqrt{\frac{B^2_{\rm m}}{\mu_0 \rho} +V^2}},
\label{equ_P2}
\end{equation}
where $R_{\rm m}$ is the moon radius, $B_{\rm m}$ is the planet 
magnetic field strength at the moon position, $V$ is the difference
between the velocity of the planet magnetosphere at the moon
position and the moon velocity
\begin{equation}
V = \omega d_{\rm m} - \sqrt{\frac{G M}{d_{\rm m}}},
\end{equation}
where $\omega$ is the planet's angular velocity and $M$ is the
mass of the planet. This difference depends on the distance
between the planet and the moon ($d_{\rm m}$). At some point,
the difference between the velocities reaches the maximum value. In 
the Jupiter--Io system this maximum point is located at about 5.5
Jupiter's radii. Note that the Io's orbit has radius slightly
larger ($\sim6 \; R_{\rm J}$). Since the radiated power is 
proportional to the square of $V$, the power radiated in the
Jupiter--Io system is almost maximal. In our estimations we 
always assume the distance $d_{\rm m}$ that gives the maximum value
of $V$. This again means that we have chosen a preferable 
physical conditions. Therefore, the fluxes we calculate are
maximal. The other important parameters in the equation (\ref{equ_P2})
are the plasma density ($\rho$) and the efficiency of the emission 
process ($\beta$). The plasma density is an unknown, free parameter. 
We assume that the value of this parameter can be of the order of 
plasma density around Io ($\sim 7 \times 10^{-17} \; \rm kg\: m^{-3}$). 
We also assume that the efficiency is similar like in the 
Jupiter--Io system $\beta=0.01$ ($\sim 1\%$). Another free
parameter is the planet's angular velocity. 
In principle the lower limit for this parameter can be established at 
the critical velocity. This is the velocity at which the dynamo process is saturated. 
However, in practice the critical velocity is very low and do not give a useful constrain 
\citep[see the discussion in][]{Reiners10}. On the other hand, the upper limit for 
the angular velocity is given by the maximum angular velocity allowed by the centrifugal 
stability $\omega_{\rm max} = \sqrt{G M /R^3}$.

\begin{figure}
\centering
\includegraphics[width=\columnwidth]{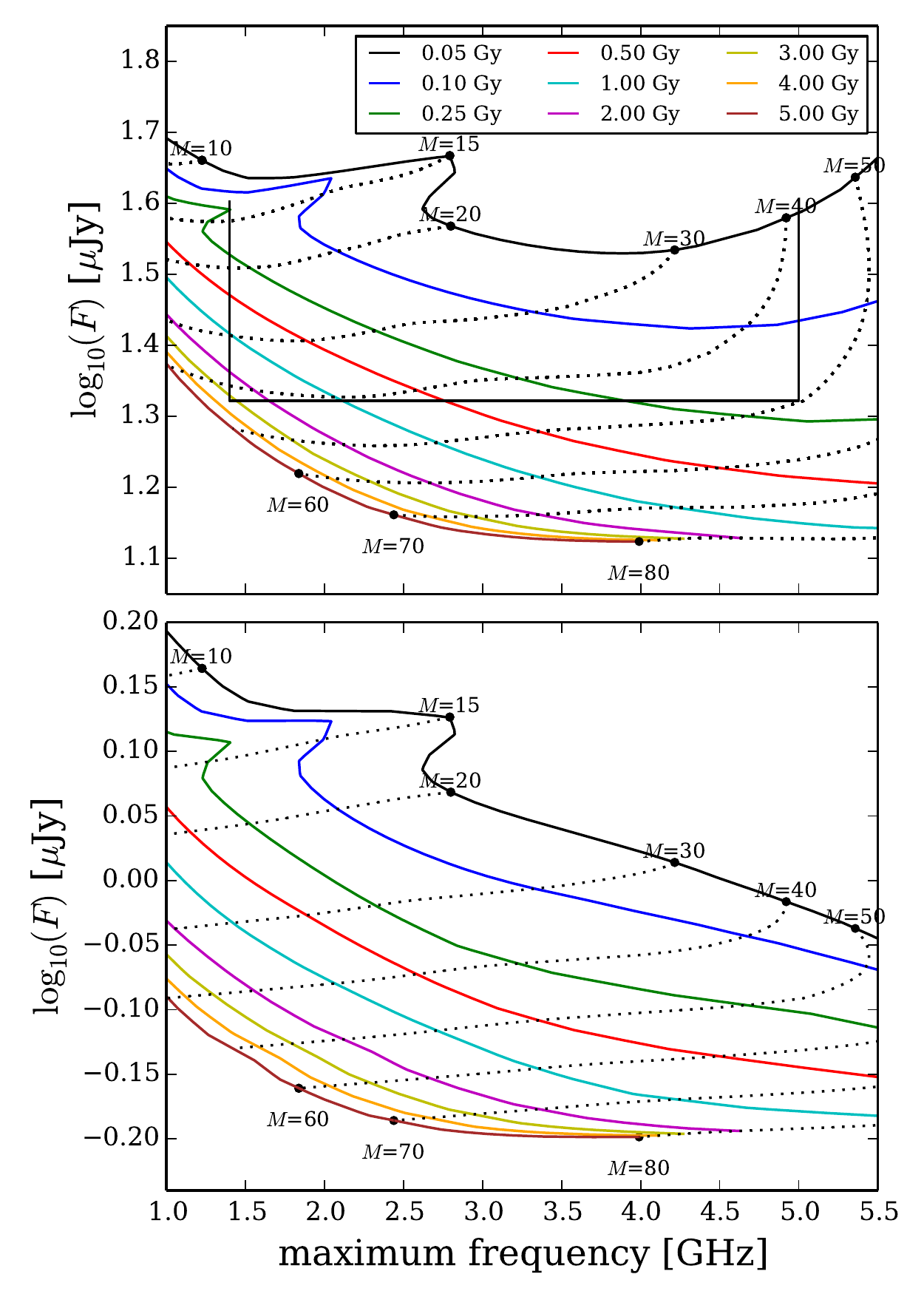}
\caption{The expected flux densities of objects located at the distance of 25 pc, generated by
the stellar wind (upper panel, according to Section \ref{sec_wind}) and the planet--moon interaction 
(lower panel, according to Section \ref{sec_moon}).
The continuous lines show expected emission levels for different masses and ages. 
The dotted lines join the fluxes generated by object with the 
same mass at different ages. The masses are given in the Jupiter units. 
The continuous "U" like curve shows the frequency range (1.4 -- 5 GHz) and
the $3\sigma$ detection threshold ($\sim 21 \; \mu \rm Jy$) that may be reached 
after five hours of the EVN integration with 1GB recording rate 
(see also Tabe \ref{tab_1} for other interferometric networks and recording 
rates). 
}
\label{fig_2}
\end{figure}

\section{Comparison of the emission scenarios}

The observed flux densities from both above discussed processes are calculated 
according to the formula
\begin{equation}
F = \frac{P_{1/2}}{\Omega D^2 \nu_{\rm cyc}},
\label{eq_flux}
\end{equation}
where $\Omega = 1.6$ sr is the solid angle of the emission beam \citep{Zarka04}, 
$D$ is distance to the planetary system and $\nu_{\rm cyc} = e B/(2 \pi m_e)$
is the cyclotron frequency.

The results of our calculations are illustrated in Fig.~\ref{fig_2}. In the
upper panel, we show the emission generated by the interaction between the
stellar wind and the planet magnetosphere. To calculate this emission we 
assumed the density of the wind equal to the density in the Solar system
$n = n_{\rm 1au}$ and the wind velocity three times higher than in the Solar system
$v = 3 v_{\rm 1au}$. Our analysis presented in the next section shows that
such numbers appears to be an average values for the main sequence A--type stars,
that we analyse in this work. Moreover, we assumed the planet to star distance 
to be $d = 1 \rm au$ and we located our hypothetical planetary systems at the 
distance $D=25\rm pc$. The first distance was assumed arbitrary, the second one
is typical for the stars analysed in the next section. Note that the formulas
that describe the emitted power and  the observed flux density are simple power-law 
functions. Therefore one can easily recalculate our results for other distances 
($d$, $D$). The first panel in Fig.~\ref{fig_2} shows that the emission of almost 
all young system (age $ \lesssim 1 $Gyr) should be detected by the EVN after five hours of 
integration, if the level of this emission will remain constant. Note that such 
emission very likely will be variable with the periods of higher activity and 
possible outburst. It is difficult to predict exact character of such emission 
for extrasolar planets. Therefore, the effective time required for the detection 
can be significantly longer.

\begin{table}
\setlength{\tabcolsep}{2.5pt}
\begin{tabular}{lrrrrrr}
Name & $\frac{M}{M_{\rm J}}$ & $\frac{\rm Sm. axis}{\rm au}$ & $\frac{D}{\rm pc}$ & $\frac{M_*}{M_{\odot}}$,& $\frac{\rm Age}{\rm Myr}$ & $\frac{\rm Spec.}{\rm type}$\\
\hline
$\alpha$ PsA b& 3.0$_{-3.0}^{+0.0}$ & 115.0 & 7.7 & 1.92 &  440 & A4V \\
$\beta$ Pic b & 7.0$_{-3.0}^{+4.0}$ & 9.04 & 19.44 & 1.8 &   21 & A6V \\
HR 8799 b & 7.0$_{-2.0}^{+4.0}$ & 68.0 & 39.4 & 1.56     &   60 & A5V \\
HR 8799 c & 10.0 $\pm 3.0$ & 42.9 & 39.4 & 1.56          &   60 & A5V \\
HR 8799 d & 10.0 $\pm 3.0$ & 27.0 & 39.4 & 1.56          &   60 & A5V \\
HR 8799 e & 9.0 $\pm 4.0$ & 14.5 & 39.4 & 1.56           &   60 & A5V \\
HD 95086 b & 5.0 $\pm 2.0$ & 61.5 & 90.4 & 1.6           &   17 & A8$\pm1^{(1)}$ \\
WASP-33 b & 2.1 $\pm 0.1$ & 0.026 & 116.0 & 1.495        &   40 & A5mA8F4$^{(2)}$ \\
HIP 73990 b & 21.0$_{-5.0}^{+30.0}$ & 20.0 & 125.0 & 1.72 &  15 & A9V \\
HIP 73990 c & 22.0$_{-6.0}^{+35.0}$ & 32.0 & 125.0 & 1.72 &  15 & A9V \\
HD169142 b & 30.0 $\pm 2.0$ & 22.7 & 145.0 & 1.65        &   60 & A7V \\
\hline
\end{tabular}
\caption{Planets and brown dwarfs discovered around main sequence A--type stars so far
\citep[after \url{exoplanets.eu,}][]{Schneider11}. Note that only two first object are located at 
relatively small distance and therefore were selected for our detailed investigations.
(1) -- according to \citet{Meshkat13} this is pre-main sequence star. (2) -- spectral type 
derived by \citet{Grenier99}.
}
\label{tab_2}
\end{table}

The hypothetical emission produced by the planet-moon interaction
is demonstrated in the lower panel of Fig.~\ref{fig_2}. This emission
in general appears to be significantly weaker than in the previously 
analysed case. The emission of a system similar to Jupiter-Io
(with magnetic field $> 1 \rm kG$) cannot be detected by currently 
operating VLBI networks at GHz frequencies. Therefore, to estimate
what we can expect from such emission scenario, we assumed significantly
higher values of the main parameters ($R_{\rm m} = 3 R_{\rm Io}$, 
$\rho = 3 \rho_{\rm Io}$, $\omega = 5 \omega_{\rm J}$). However, 
even with this assumption the fluxes calculated for the distance 
of 25 pc are very small ($F \lesssim 1 \mu \rm Jy$). At smaller 
distance, for example 10 pc, the number of possible targets is reduced
just to three stars, and the expected emission from such 
systems is below 10 $\mu \rm Jy$. Thus, below the sensitivity 
threshold of currently operating interferometers.
Therefore, in our further investigations we will focus 
on the first emission scenario, where the radiation is produced by the 
interaction between the stellar wind and the planet magnetosphere.

\begin{figure}
\centering
\includegraphics[width=\columnwidth]{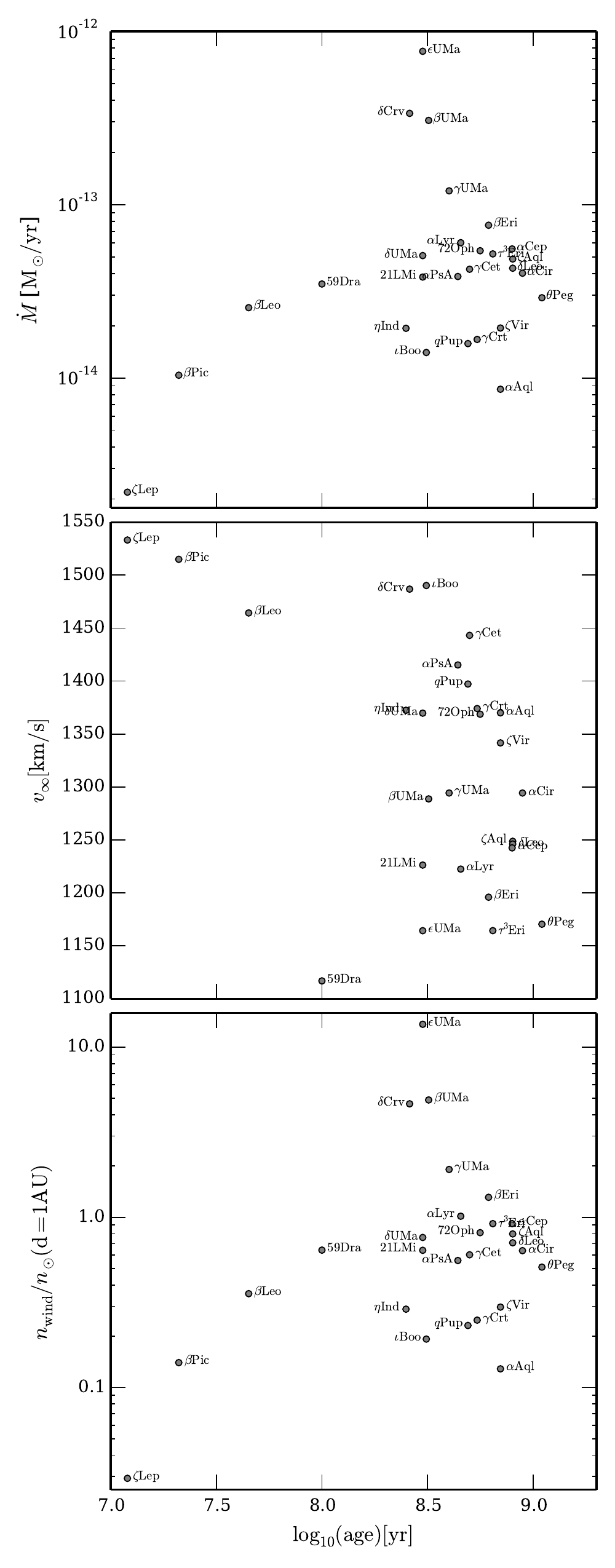}
\caption{Stellar wind parameters estimated for the sample of main sequence A--type stars
analysed in this work. The upper panel shows mass loss rates, in the middle panel we plotted
terminal velocities and the bottom panel shows wind particle densities normalized to the
Sun wind particle density observed at the distance of 1au.}
\label{fig_3}
\end{figure}

\section{Application to known stars}

\begin{table*}
\begin{center}
\begin{tabular}{lrlrrrrrc}
Name & $\frac{D^{(1)}}{\rm pc}$ & $\frac{M_*}{M_\odot}$ & $\frac{L_*}{L_\odot}\left(\frac{T_{\rm eff}}{\rm K}\right)$ & $\frac{R_*}{R_\odot}$ & $(M/H)$ & $\frac{\rm Age}{\rm Myr}$ & Spectral type & $\frac{\rm IR}{\rm exces}$\\[0.1cm]
\hline\\[-1.8ex]
       $\alpha$ Aql &  5.13 & 1.77$^{(4)}$  &  10.60$^{(38)}$        &  1.82$^{(15)}$ & -0.20$^{(34)}$ & 700$^{(10)}$ &             A7V &  y$^{(15)}$ \\
       $\alpha$ Lyr &  7.68 & 2.13$^{(6)}$  &  40.12$^{(6)}$         &  2.75$^{(26)}$ & -0.43$^{(33)}$ & 454$^{(6)}$ &            A0Va &  y$^{(7)}$ \\
       $\alpha$ PsA &  7.70 & 1.92$^{(12)}$ &  16.53$^{(12)}$        &  1.85$^{(30)}$ &  0.10$^{(32)}$ & 440$^{(12)}$ &            A4V &  y$^{(8)}$ \\
        $\beta$ Leo & 11.00 & 1.90$^{(2)}$  &  15.00$^{(30)}$        &  1.71$^{(30)}$ &  0.00$^{(30)}$ &  45$^{(2)}$ &            A3Va &  y$^{(8)}$ \\
       $\alpha$ Cep & 15.04 & 2.00$^{(24)}$ &  20.50 (7773)$^{(33)}$ &  2.50$^{(24)}$ &  0.09$^{(33)}$ & 795$^{(24)}$  &            A8Vn &  -- \\
       $\alpha$ Cir & 16.57 & 1.71$^{(21)}$ &  11.82 (7631)$^{(32)}$ &  1.97$^{(22)}$ &  0.36$^{(32)}$ & 890$^{(21)}$  &     A7Vp-SrCrEu &  y$^{(42)}$ \\
       $\delta$ Leo & 17.91 & 2.06$^{(19)}$ &  24.97$^{(19)}$        &  2.56$^{(19)}$ & -0.18$^{(31)}$ & 800$^{(19)}$ &           A5IVn &  -- \\
        $\beta$ Pic & 19.44 & 1.76$^{(4)}$  &   8.70$^{(37)}$        &  1.48$^{(20)}$ &  0.05$^{(32)}$ &  21$^{(13)}$ &             A6V &  y$^{(14)}$ \\
        $\zeta$ Lep & 21.61 & 1.90$^{(2)}$  &  10.56 (8337)$^{(32)}$ &  1.56$^{(15)}$ & -0.76$^{(32)}$ &  12$^{(2)}$ &       A2IV-V(n) &  y$^{(8)}$ \\
        $\zeta$ Vir & 22.71 & 1.94$^{(19)}$ &  17.89$^{(19)}$        &  2.08$^{(19)}$ & -0.26$^{(31)}$ & 700$^{(19)}$ &           A2Van &  -- \\
         $\eta$ Ind & 24.17 & 1.62$^{(25)}$ &  7.62  (7448)$^{(32)}$ &  1.66$^{(20)}$ &  0.40$^{(32)}$& 250$^{(25)}$  &         A9IV &  y$^{(43)}$ \\
       $\gamma$ Cet & 24.41 & 2.18$^{(4)}$  &  20.74 (8673)$^{(33)}$ &  2.02$^{(28)}$ &  0.00$^{(33)}$ & 500$^{(10)}$ &            A2Vn &  -- \\
        $\beta$ UMa & 24.45 & 2.60$^{(2)}$  &  63.01$^{(19)}$        &  3.02$^{(19)}$ & -0.03$^{(31)}$ & 320$^{(2)}$ &          A1IVps &  y$^{(8)}$ \\
       $\delta$ UMa & 24.69 & 2.10$^{(2)}$  &  23.07 (8613)$^{(33)}$ &  2.16$^{(15)}$ & -0.03$^{(33)}$ & 300$^{(2)}$ &            A2Vn &  y$^{(11)}$ \\
       $\gamma$ Crt & 25.24 & 1.80$^{(4)}$ &  11.30 (7805)$^{(32)}$ &  1.84$^{(18)}$ &  0.04$^{(32)}$ & 543$^{(4)}$  &          A7Vn &  y$^{(5)}$ \\
     $\epsilon$ UMa & 25.31 & 2.91$^{(9)}$  & 101.92 (9020)$^{(33)}$ &  4.14$^{(27)}$ &  0.00$^{(33)}$ & 300$^{(10)}$ &      A1III-IVp &  -- \\
        $\zeta$ Aql & 25.46 & 1.98$^{(19)}$ &  38.49$^{(19)}$        &  2.45$^{(19)}$ & -0.52$^{(31)}$ & 800$^{(19)}$ &        A0IV-Vnn &  y$^{(3)}$ \\
       $\gamma$ UMa & 25.50 & 2.64$^{(4)}$  &  63.75 (9361)$^{(31)}$ &  3.04$^{(26)}$ & -0.44$^{(31)}$ & 400$^{(5)}$ &            A0Ve &  -- \\
       $\delta$ Crv & 26.63 & 2.59$^{(4)}$  &  69.00$^{(36)}$        &  2.26$^{(20)}$ & -0.07$^{(32)}$ & 260$^{(5)}$ &           A0IVn &  y$^{(20)}$ \\
             72 Oph & 26.63 & 1.99$^{(4)}$ &  17.80 (8400)$^{(33)}$ &  2.05$^{(20)}$ &  0.20$^{(33)}$ & 561$^{(4)}$  &            A4IV &  -- \\
     $\tau^3\!$ Eri & 27.17 & 1.89$^{(2)}$  &  27.23 (8045)$^{(32)}$ &  2.69$^{(29)}$ & -0.21$^{(32)}$ & 644$^{(2)}$ &          A3IV-V &  -- \\
             59 Dra & 27.30 & 1.70$^{(17)}$ &  17.19 (7252)$^{(39)}$ &  2.63$^{(23)}$ & -0.03$^{(40)}$ & 100$^{(10)}$ &            A9V &  -- \\
        $\beta$ Eri & 27.40 & 2.32$^{(18)}$ &  36.08 (8002)$^{(31)}$ &  3.13$^{(18)}$ & -0.20$^{(31)}$ & 615$^{(18)}$ &           A3III &  y$^{(41)}$ \\
             21 LMi & 28.24 & 1.80$^{(2)}$  &  18.10 (7839)$^{(33)}$ &  2.31$^{(29)}$ & -0.01$^{(33)}$ & 300$^{(2)}$ &            A7Vn &  y$^{(17)}$ \\
       $\theta$ Peg & 28.25 & 1.86$^{(19)}$ &  24.55$^{(19)}$        &  2.62$^{(19)}$ & -0.38$^{(32)}$ & 1100$^{(19)}$ &            A2Vp & -- \\
              q Pup & 28.63 & 1.76$^{(18)}$ &  10.02 (7790)$^{(32)}$ &  1.74$^{(18)}$ &  0.11$^{(32)}$ & 491$^{(18)}$  &             A8V &  -- \\
        $\iota$ Boo & 29.07 & 1.68$^{(4)}$ &   9.55$^{(35)}$ &  1.46$^{(23)}$ &  0.08$^{(35)}$ & 312$^{(4)}$         &             A9V &  -- \\       
\hline
\end{tabular}
\end{center}
\caption{Parameters of stars selected for the calculations presented in Section 6. In the cases where the
         bolometric luminosity was not directly available, we give the effective temperature (in brackets), 
         that we used to calculate the luminosity. References:  
             1) \citet{vanLeeuwen07},     
             2) \citet{Chen14},     
             3) \citet{Chen05}, 
             4) \citet{Zorec12},          
             5) \citet{Gaspar13},
             6) \citet{Yoon10},           
             7) \citet{Aumann84},
             8) \citet{Cote87},
             9) \citet{Shaya11},         
            10) \citet{Nakajima12}
            11) \citet{Su06},
            12) \citet{Mamajek12},
            13) \citet{Binks14}, 
            14) \citet{Smith84},
            15) \citet{Absil13},
            16) \citet{Morales09},
            17) \citet{Galland06},
            18) \citet{daSilva06},
            19) \citet[A]{Boyajian12},
            20) \citet{Ertel14},
            21) \citet{Kochukhov06},
            22) \citet{Bruntt08},
            23) \citet{vanBelle09},
            24) \citet{vanBelle06},
            25) \citet{Plavchan09},
            26) \citet{Fitzpatrick05},
            27) \citet{Shulyak14},
            28) \citet[B]{Boyajian12b},
            29) \citet{vanBelle12},
            30) \citet{DiFolco04},
            31) \citet{Wu11},
            32) \citet{Gray06},
            33) \citet{Gray03},
            34) \citet{Monnier07},
            35) \citet{Paunzen02},
            36) \citet{Montesinos09},    
            37) \citet{Crifo97},
            38) \citet{Peterson06},
            39) \citet{King03},
            40) \citet{Boesgaard88},
            41) \citet{Trilling07},
            42) \citet{McDonald12},
            43) \citet{Plavchan09}.
}
\label{tab_3}
\end{table*}

The calculations presented in the previous section demonstrate that the
strongest emission could be generated in the youngest systems as old as about
a few hundreds Myr (Fig.~\ref{fig_2}), where the magnetic field strength
may reach the highest values (Fig.~\ref{fig_1}). Therefore, looking for
star candidates that may host young planets or brown dwarfs we selected 
main sequence A-type stars located in the solar neighbourhood ($D\la 30\,$pc).
 What is important, these objects evolve relatively fast and usually 
leave the main sequence in less than 1\,Gyr. Thus, possible planets around
such stars should also be relatively young. Moreover, the main sequence A-type 
stars are more massive than the Sun (masses from 1.3 to $3 M_\odot$). 
Statistical analyses of known planetary systems suggests that planets 
originate more frequently around massive stars 
\citep[e.g.,][]{Johnson07}. The observed fraction of stars
with giant planets is: $~3.3 \%$ for M-type dwarfs, $8.5 \%$ for F, G, K type
stars and $20 \%$ in the case of `retired' A-type objects \citep{Johnson10}. 
On the other hand there is a theoretical prediction that massive planets should 
originate more frequently around less massive stars \citep{Kornet06}.
However, this work is based on the core accretion theory that may have
difficulties to explain the existence of giant planets relatively close to
massive stars \citep{Ribas15}.
Moreover, there are only a few planets discovered around main sequence A-type 
stars, so far (see Table~\ref{tab_2}). However, this is related rather to difficulties
in detection of such planetary system, where for example fast rotation
of a star or less number of spectral lines rules out standard radial 
velocity measurements. From the planets listed in Table~\ref{tab_2}, one object  
\citep[WASP-33 b -- ][]{Collier10} was discovered by the observations of transits. 
The rest of these planets were detected by the direct imaging. This indicate significance 
of this technique, that may be even more important in the radio domain.

Physical parameters of main sequence A-type stars selected for the calculations
presented in this section are collected in Table~\ref{tab_3}. The errors for the most
of the specified parameters are at level of a few percent, with the exception for the age.
This parameter is the most difficult value for the estimation, especially in the case of
main sequence stars, that for relatively long period of time radiate almost the same amount 
of energy. Thus, in many papers the errors of the estimated ages are not specified at all. From
all collected informations we may conclude that the age of Vega ($454 \pm 13$ Myr) 
is estimated with the best precision (better than 3\%), whereas the age of $\eta$ Ind 
($250 \pm 200$ Myr) has the highest uncertainty (80 \%).

More than half of the selected objects 
exhibit an excess in the infrared range. This may indicate that these stars are surrounded 
by debris discs, what increases a chance to find planets or brown dwarfs in such systems. 
Note that for three from the selected stars ($\alpha$ Lyr, $\beta$ Pic, $\alpha$ PsA) debris 
discs were already directly observed \citep{Holland98}, and the massive Jovian planets were
discovered inside the debris disc of $\alpha$ PsA \citep{Kalas08} and $\beta$ Pic 
\citep{Lagrange09}. Finally, our putative planetary systems are located at relatively small 
distances, what should help to detect possible radio emission.

The most important stellar wind parameters obtained for the selected stars are
presented in Fig.~\ref{fig_3}. Expected mass--loss rates extends over two orders of 
magnitude. This is the result of differences in radii and luminosities of these
stars. An average value of the mass loss rate $<\dot{M}> = 8.6 \times 10^{-14} \; 
\rm M_\odot/yr$ is a few times higher than the Sun mass loss rate ($\dot{M}_\odot \simeq 2 
\times 10^{-14} \; \rm M_\odot/yr$). However, $\dot{M}$ of most of the stars lies
in the range between $10^{-14}$ and $10^{-13} \; \rm M_\odot/yr$, which is close
to $\dot{M}_\odot$. Estimated terminal velocities are in range 1100 -- 1550 km/s.
This is 2 to 4 times higher than the velocities observed in the Solar wind. Note
that in the Solar wind there are two components, slower and heavier with the velocity
around 425 km/s and the faster component with the velocity $\sim$750 km/s. Expected
wind particle densities for most of the stars are in the range $ 0.1 < n < 10$ in
comparison to the Sun wind particle density at the distance of 1au. However, an
average value of this parameter is similar to the value observed in the Solar
system ($n_{\rm 1au} \simeq 6.6 \times 10^6 \rm m^{-3}$).

\begin{figure}
\centering
\includegraphics[width=\columnwidth]{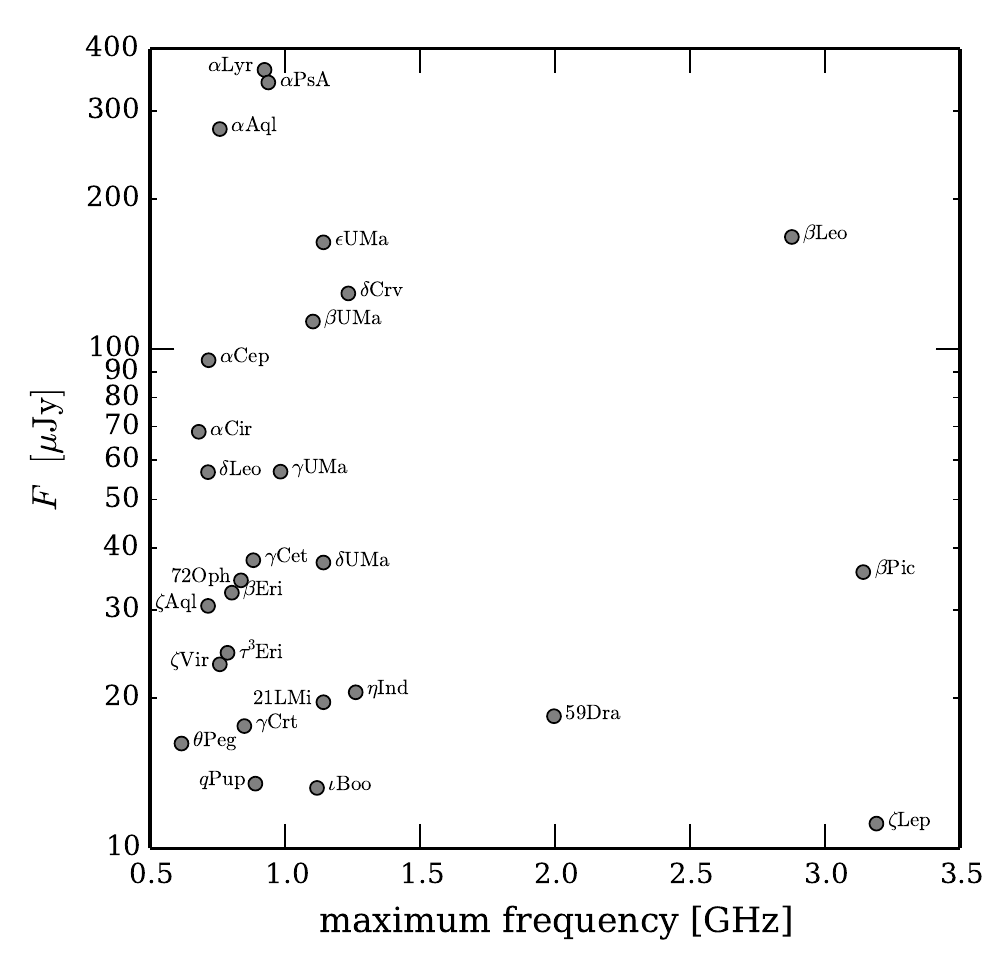}
\caption{
The expected radio emission from hypothetical object with masses $M=15 M_{\rm J}$ 
located at the distance $d= 1\rm au$ around selected main sequence A-type stars. 
}
\label{fig_4}
\end{figure}

In Fig. \ref{fig_4} we show possible radio emission that can be generated
by the interaction between the stellar wind and objects orbiting around selected stars. 
To calculate expected maximal frequencies and fluxes we assumed that in each system 
there is an object with the mass $M=15 M_{\rm J}$, located at the distance 
$d= 1 \rm au$ from the star. 
For most of the selected stars the maximum frequency of the emission appears around 
1 GHz and the expected flux lies in the range from 10 to 400 $\mu$Jy.

Our calculations show that the strongest radio emission should be expected from
hypothetical planetary system around Vega ($\alpha$ Lyr) and Fomalhaut ($\alpha$ PsA). 
These two stars are located at the distance $\sim$7.7 pc. Thus, the observed flux 
density should be relatively high $\sim 350\mu \rm Jy$.
The maximum frequency of the radio emission in these systems may reach $\sim$0.9~GHz. 
Note that \citet{Brown90} using VLA made observations of several A and F type 
stars at 4835 and 4885 MHz (50 MHz bandwidth). However, they obtained no detections,
only upper limits were estimated. In the case of Fomalhaut the upper limit was set
to $F_{4.8 \rm GHz} \leq 100 \mu \rm Jy$. As it was already mentioned Vega is surrounded by the debris 
disc \citep{Holland98}. The close neighbourhood of this star may contain also dust 
\citep{Absil06}. The direct imaging of the disc around Fomalhaut led to the detection 
of a planet \citep{Kalas08}. This planet is located at the distance of at least 100 au from 
the star \citep{Kalas13}. Therefore, the emission produced by the stellar wind 
should be negligible. On the other hand, if there is a moon around this planet that through volcanic 
activity fills the planet magnetosphere with matter, then we may expect 
the radio emission from such system. However, as we already demonstrated this 
emission process is relatively weak at GHz frequencies. Thus, from the 
distance of almost 8 pc it would be difficult to detect such emission. 
Still, detection of a radio emission from a planet located tens of astronomical
units from the star can indicate an existence of extrasolar moons.

The emission similar to that of Vega and Fomalhaut can also be expected 
from Altair ($\alpha$ Aql). This star is located at the distance $\sim$5 pc. 
Therefore, the expected flux density is also relatively high $\sim$280$\; \mu \rm Jy$. 
However, this is quite old object (age $\sim$700 Myr). Therefore, the magnetic 
field strength in this system may be too weak to produce sufficient emission 
at GHz frequencies (maximum emission frequency is expected to be around 760 MHz). 
According to \citet{Brown90} the upper limit for the emission in this system 
is $F_{4.8 \rm GHz} \leq 100 \mu \rm Jy$. On the other hand this can be an 
interesting target for the low frequency interferometers (GMRT \& LOFAR).

Another interesting star is $\beta$ Pictoris. The planet discovered around this
star \citep{Lagrange09} is massive (4-11 $M_{\rm J}$) and located at relatively
small distance to the star $d \sim1.5d_{\rm J}$ \citep{currie13}. The system is
also very young (age $\sim 21$ Myr), therefore it fulfils all the necessary conditions
to be detected at the radio GHz frequencies. Unfortunately, the star is rather
distant ($D = 19.4$ pc) and located at the Southern hemispere ($\delta \simeq 
-51^\circ$), beyond the operating range of the most sensitive interferometers 
(VLBA, EVN). Our calculations show that the flux density of the radio emission
from the planet in this system should be at the level of about 30 $\mu$Jy above
3~GHz. Thus, this is an excellent target for the Square Kilometre 
Array\footnote{\url{http://www.skatelescope.org/}} in near future.

The most promising candidate for detection of the radio emission at GHz 
frequencies from massive exoplanet or a brown dwarf is Denebola ($\beta$ Leo). 
This star is young ($\sim$45~Myr) and located at relatively small distance ($\sim$11~pc). 
Therefore, a massive object ($M=15 M_{\rm J}$) in this system, located at the distance of 1 au
should generate the flux around 166 $\mu$Jy at $\sim 2.9$ GHz. It must be mentioned that
the upper limit obtained by \citet{Brown90} for this star is $F_{4.8\,\rm GHz} \leq \; 110\mu$Jy. 
However, this does not exclude the possibility that planets are present around this star.
Our calculations show that Denebola is the best candidate for search of possible
planetary emission at GHz frequencies.

\section{Summary}

We calculated possible radio GHz emission that may originate in
extrasolar planetary systems. The key parameter for the frequency
range and the power of such emission is the magnetic field strength.
This parameter was obtained from the theoretical model that describes
the evolution of massive planets and brown dwarfs. We analysed two emission 
scenarios that are observed in the Jupiter system. The first mechanism
gives quite promising results. This emission could be detected from
massive planets ($M \ga 10 M_{\rm J}$) and brown dwarfs at 
the distances $\la 30$ pc. The less optimistic is the fact that the 
emission from planets is limited to young and massive objects and could 
be observed practically only at 1.4~GHz by global VLBI systems. Using these 
results we selected several young and massive A-type stars to calculate possible 
radio emission from planets and brown dwarfs around such objects. Our 
estimations demonstrated that in almost all cases the emission can be 
detected by VLBI.

The interaction between hypothetical moons and planets that may produce
some radio emission appears to be less significant. In principle, the observations 
of such emission may give a direct evidence for an existence of extrasolar moons 
\citep{Noyola14}. However, such a planet with the moon should be located at relatively 
large distance from the star (at least several au) to exclude possibility that the 
radio emission is dominated by the interaction of the star wind with the planet 
magnetosphere. Our calculations show that such detection can be very difficult 
with currently operating instruments.

\section*{Acknowledgements}
We thank the anonymous referee for very detailed report that helped us
to significantly extend and improve this manuscript. We thank also to 
Roman Schreiber for helpful comments and discussions. We are grateful 
to Polish National Science Centre for their support of the RISARD project 
(grant no. 2011/01/D/ST9/00735). This research has made use of the SIMBAD 
data base operated at CDS, Strasbourg, France.

\bibliographystyle{mnras} 
\bibliography{ms}

\bsp

\label{lastpage}

\end{document}